\documentclass [12pt,amsmath, amssymb, aps, bbm, nofootinbib, noindent, notitlepage] {article}
 \usepackage{lipsum}
\usepackage[margin=1.75cm]{geometry}
\usepackage{amssymb}
\usepackage{graphicx}
\usepackage{amsmath}
\usepackage[hidelinks]{hyperref}
\usepackage{array}
\usepackage{verbatim} 
\usepackage{multirow}
\usepackage{manfnt}
\usepackage{mathrsfs}
\usepackage{hyperref}
\usepackage{float}
\usepackage{caption}
\usepackage{braket}
\usepackage[numbers]{natbib}
\usepackage{authblk}
 \usepackage{lipsum}

\usepackage{csquotes}

\numberwithin{equation}{section}

\usepackage{tocloft}

\usepackage{tocloft}

\def\be{\begin{equation}}
\def\ee{\end{equation}}

\newcommand{\SU}{\mathrm{SU}(2)}

\usepackage{authblk}
\usepackage{color}

\begin{document}
\title{Quantum information elements in Quantum Gravity states and processes}
\author{ Daniele Oriti \thanks{doriti@ucm.es}}
\affil{\small Depto. de Física Teórica and IPARCOS, Facultad de Ciencias Físicas \\
Universidad Complutense de Madrid, Spain, EU}

\maketitle

\abstract{We summarize basic features of quantum gravity states and processes, common to a number of related quantum gravity formalisms, and sharing a purely combinatorial and algebraic language, and a discrete geometric interpretation. We emphasize how, in this context, entanglement is a seed of topological and geometric properties, and how a pre-geometric, discrete notion of quantum causality can be implemented, as well as some recent results (based on random tensor network techniques) on the conditions for information transmission and holographic behaviour in quantum gravity states. Together, these features  indicate that quantum information concepts and tools play a key role in defining the fundamental structure of quantum spacetime.  
 }



\section*{Introduction}
The main goal of this contribution is to show how quantum gravity states and processes as identified by a number of quantum gravity formalisms can be recast in the language of quantum information and how entanglement or quantum correlations can then be seen, in the same formalisms, as essential in the very structure of quantum spacetime. It is not a review even of the few results we will summarize briefly, let alone of the substantial research done on entanglement and quantum information features in quantum gravity formalisms. For the latter, we refer to \cite{Bianchi:2023avf,Colafranceschi:2022ktj}, to remain limited to the results obtained in the quantum gravity context closer to our focus.

The perspective we find convenient to adopt, in order to appreciate the role of quantum information-theoretic structures in these quantum gravity formalisms, is that of emergent spacetime, i.e. of quantum gravity as a theory of \lq spacetime constituents\rq with spacetime itself, geometry and fields as emergent entities \cite{Butterfield:1999ah, Seiberg:2006wf,Padmanabhan:2014jta, Oriti:2018dsg, Carlip:2012wa}.
 This perspective is motivated by several results in semiclassical physics, for example black hole thermodynamics and  the information paradox, gravitational singularities, that all point in various ways to a breakdown of key notions on which standard continuum, geometric physics is based, and, more indirectly, the results of analogue gravity in condensed matter systems, showing how effective field theory on curved backgrounds can emerge rather generically from non-gravitational systems. It is also motivated by results in modern quantum gravity approaches, including the ones we focus on in this contribution, with their combinatorial, algebraic and, indeed, (quantum) information-theoretic structures replacing geometric notions and spacetime-based quantum fields. Indeed, to give some examples, canonical Loop Quantum gravity replaces smooth metric geometries with piecewise-degenerate quantum twisted geometries encoded in combinatorial/algebraic data, lattice quantum gravity works with piecewise-flat quantum (often non-metric) geometries, string theory dualities also suggest that the fundamental degrees of freedom of M-theory are not spacetime-based, and AdS/CFT gives a concrete example of emergent gravity as well as a very partial \lq emergent space\rq, reconstructed from a lower-dimensional and non-gravitational CFT.
 This perspective also implies a shift away from the more traditional perspective that sees quantum gravity as the result of straightforwardly quantizing General Relativity or some other classical gravitational and spacetime-based theory (whether perturbatively or non-perturbatively)\footnote{Interestingly, this more traditional perspective is also, historically and in part of the present community, the one adopted to interpret some of the same quantum gravity formalisms that we suggest can be fruitfully understood from an emergent spacetime perspective.}. 
 From an emergent spacetime perspective, a breakdown of spacetime notions, including locality, should be expected when moving to a more fundamental description. One the key task is then to identify the hidden, possibly discrete microstructure replacing continuum spacetime fields in such more fundamental description of the universe, with such fields, including the metric, being then understood as collective entities and gravity and the rest of continuum spatiotemporal physics as an approximate effective description of collective dynamics; controlling such collective dynamics is the second key task. In other words, the universe itself is seen as a (background independent) quantum many-body system. 

Tied to the notion of spacetime emergence is the further hypothesis that spacetime geometric and, possibly, topological structures can in fact emerge from the entanglement among more fundamental quantum constituents \cite{VanRaamsdonk:2016exw, Swingle:2017blx}, thus via a conjectured \lq entanglement/geometry correspondence\rq.  Support for this conjecture has been obtained mostly in a semi-classical context and in the AdS/CFT context, thus in presence of well-defined spacetime and geometric notions, starting with the Ryu-Takanayagi entropy formula and related results \cite{Rangamani:2016dms}. However, they are suggestive of something more fundamental, that calls for a concrete realization of this idea in full quantum gravity, thus also in absence of spacetime and fields as we know them. This call has been heard and partially answered, we would claim, in the quantum gravity formalisms on which the rest of this contribution will focus. Quantum correlations and, more generally, quantum information-theoretic notions acquire, in fact, a central role.
These formalisms are canonical loop quantum gravity, spin foam models, group field theories and lattice quantum gravity in first order (tetrad-connection) variables. Despite several technical differences between them, they all share many basic features. We will discuss such shared features, and only occasionally point out specific differences; unless noted otherwise, we will only consider models of 4-dimensional quantum gravity and spacetime and a Lorentzian (as opposed to Euclidean/Riemannian) setting.
We will first discuss the nature of quantum gravity states in these formalisms, and emphasize the role that entanglement among their constituents plays, their re-interpretation as quantum circuits and their use to define holographic maps and quantum information channels.
Then, we will discuss the corresponding quantum gravity processes, indicating their possible formulation as quantum causal histories and, again, as quantum circuits, as well as the present limitations to such reformulation.

\section{Quantum gravity states as entanglement networks and quantum circuits}
In the quantum gravity formalisms we are concerned with here, generic quantum gravity states can be represented as (superpositions of) entanglement networks of quantum geometric constituents; more precisely, they are expressed by assigning algebraic data to a combinatorial graph, where the algebraic data are taken from the (representation) theory of (Lie) groups, notably the Lorentz group or the rotation subgroup thereof. In turn, they can be seen as composed of elementary quantum systems, with associated one-body Hilbert space, located on nodes of the graph, with the graph itself encoding a pattern of entanglement across the (sub-)systems living on the nodes.

\begin{figure}[!ht]
\centering
\includegraphics[width=0.5\linewidth]{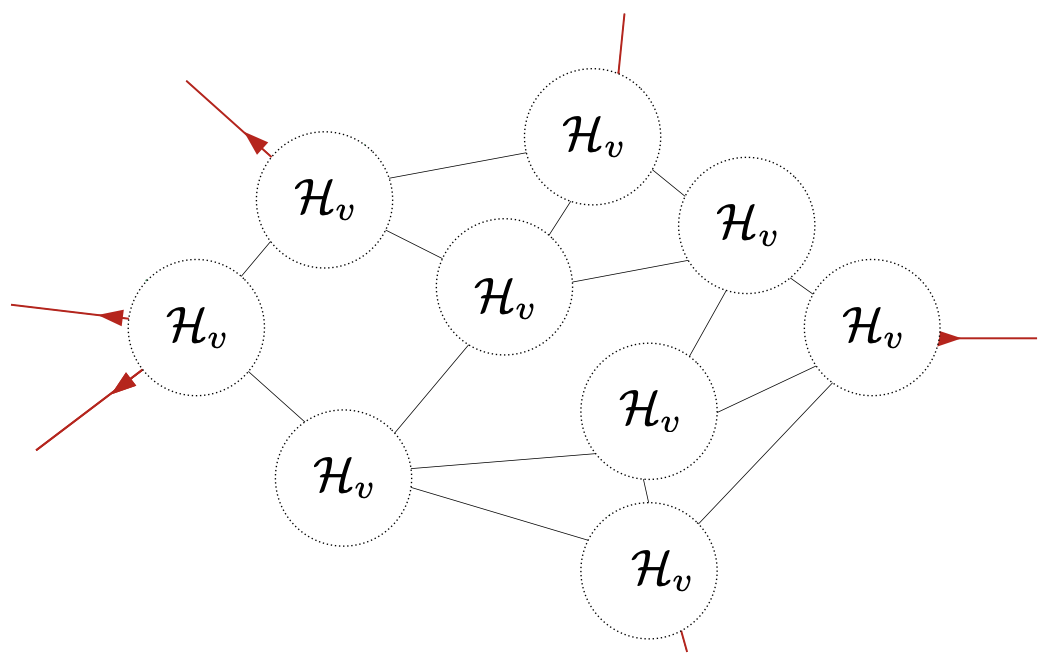}
\caption{\label{fig:frog}The general structure of quantum gravity states as many-body systems., with the underlying graph including both open links and links connecting pairs of nodes.}
\end{figure}

The graphs are usually taken to be dual to 3-dimensional simplicial complexes, in quantum gravity formalisms with a direct discrete geometric interpretation. In the following we restrict to such case. Different formalisms, as well as different models within a given formalism, will differ for the class of graphs being considered and for the choice of one-body Hilbert space associated to the nodes of the graphs, as well as for the quantum dynamics assumed for such quantum gravity states. Leaving dynamical considerations aside, in this section, we now give more details on a specific (set of) proposal(s) for the kinematical structures, and illustrate their quantum information-theoretic aspect.

\subsection{Entanglement patterns of atoms of space - quantum space as a quantum circuit}

The one-body Hilbert space can be taken to be space of states for a quantum tetrahedron, which can be constructed from $\SU(2)$ group data and expressed in terms of its irreducible representations:

\begin{equation}
\mathcal{H}_v = \bigoplus_{{\vec{j}_v}}\left( \otimes_{i=1}^{4} V^{j_v^i} \bigotimes \mathcal{I}^{\vec{j}_v}\right)
\label{Hilbert-tetra} 
\end{equation}
where one has a vector space $V^{j_v^i}$ for the representation label $j_i$ (a \lq spin\rq valued in the half-integers) for each of the four triangles of the tetrahedron, with canonical basis $| j^i , n^i\rangle$, then tensored together, and $\mathcal{I}^{\vec{j}_v}= Inv_{G}[ V^{j_v^1} \otimes \cdots\otimes V^{j_v^4}]$ is the space of intertwiners, i.e. tensors invariant under the diagonal action of the group $G=\SU(2)$, built from the same four representation spaces.
This Hilbert space can be obtained from the direct quantization of the classical phase space of geometries of a single tetrahedron, parametrized by Lie algebra elements corresponding to normal vectors associated to its four triangles and conjugate group elements corresponding to elementary parallel transports of an $\SU(2)$ connection along paths dual to the same triangles. It can depicted dually as a single vertex with four semi-links outgoing from it (each dual to one of the four triangles of the tetrahedron). 

\begin{figure}[!ht]
\centering
\includegraphics[width=0.3\linewidth]{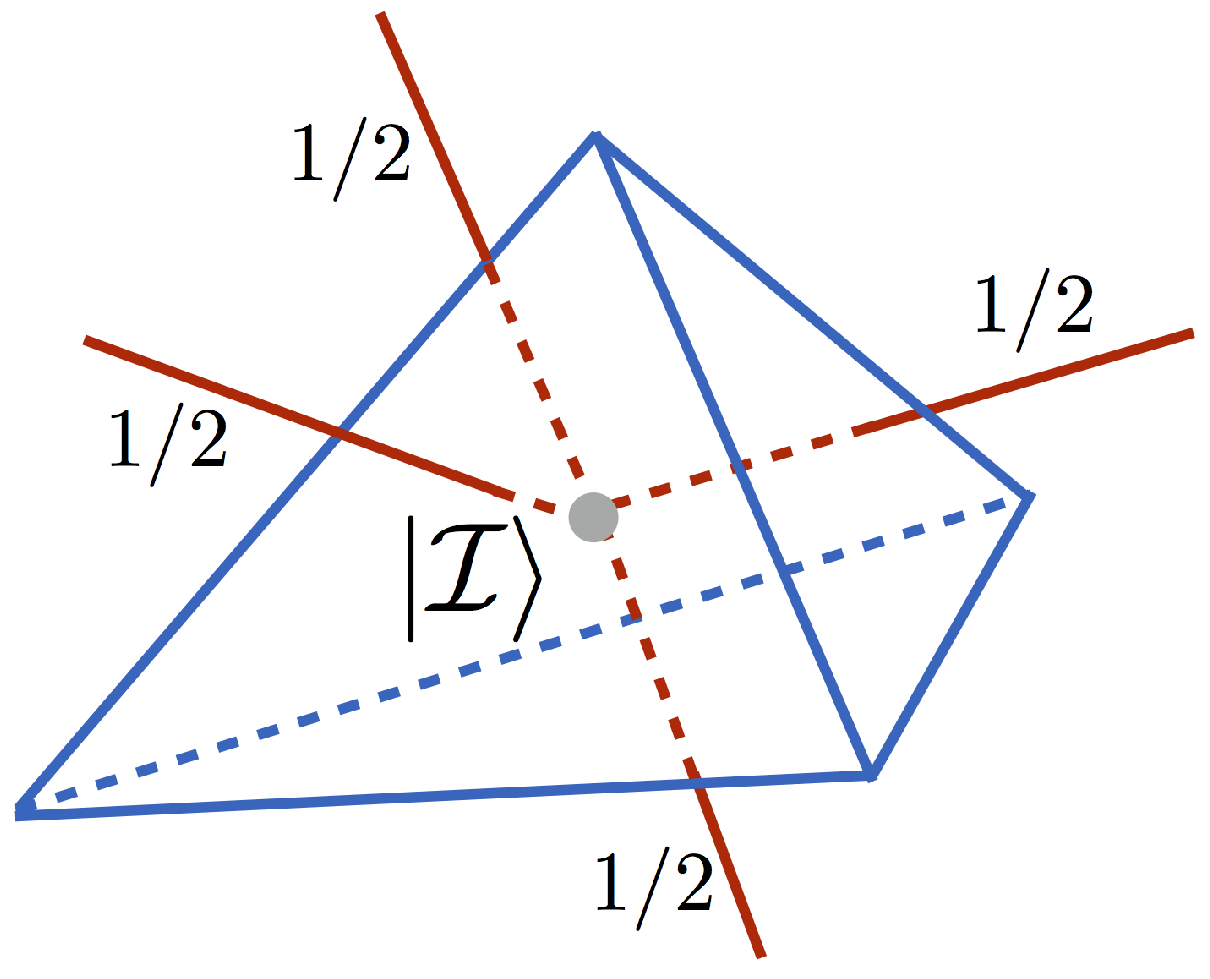}
\caption{\label{fig:SN-TET} \small A spin network vertex, with specific choice of labels, dual to a 3-simplex (tetrahedron).}
\end{figure}

The discrete geometric interpretation is confirmed by the action of geometric operators encoding the tetrahedral geometries. For example, elements of the canonical basis in $V^{j_v^i}$ diagonalize the area of the corresponding triangle, while intertwiners encode information about the volume of the whole tetrahedron with given triangle areas. For more details about this quantum geometry, see \cite{Pereira:2010wzm, Perez:2012wv}.
Thus, generic states in the (kinematical) Hilbert space of the quantum gravity formalisms we consider here can be understood as quantum many-body states built out of this single-body Hilbert space.

In particular, an interesting class of quantum states are those that admit a natural interpretation as corresponding to quantum tetrahedra glued to one another across shared faces to form extended simplicial complexes dual to 4-valent graphs. These are (maximally) entangled states of quantum tetrahedra \cite{Baytas:2018wjd,Colafranceschi:2020ern}, and can be obtained by imposing, on a state corresponding to a set of N disconnected quantum tetrahedra $| \psi\rangle \in \mathcal{H}_N = \bigotimes_{n=1}^{N} \mathcal{H}^{n}_v$, the projector $P_\gamma = \prod_{A_{xy}^i=1} P_{i}^{xy}$, where $A_{xy}^i$ is the adjacency matrix of the graph $\gamma$ ($(x,y)$ label the pairs of vertices in the graph, and the additional index $i$ runs trough the possible multiple links connecting the same two vertices), and the gluing projector $P_i^{xy} : \mathcal{H}_i^x \otimes  \mathcal{H}_i^y \rightarrow Inv\left( \mathcal{H}_i^x \otimes  \mathcal{H}_i^y\right)$ imposes maximal entanglement along the corresponding degrees of freedom of the two semi-links one intends to connect, by tracing over the corresponding $SU(2)$ labels, and thus imposing (diagonal) $SU(2)$ invariance. 

\begin{figure}[!ht]
\centering
\includegraphics[width=0.4\linewidth]{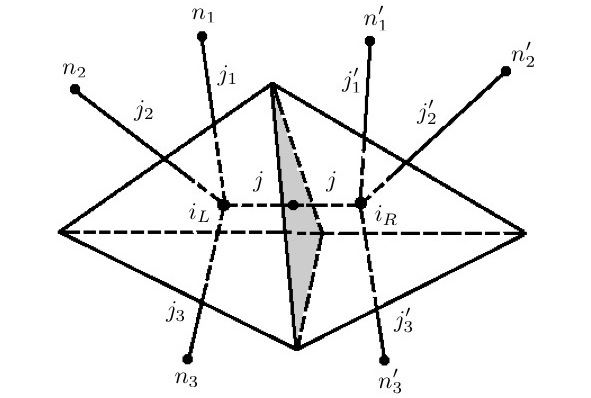}
\caption{\small The gluing of two spin network vertices (equivalently, two quantum tetrahedra) obtained by the imposition of maximal entanglement across the shared semi-link degrees of freedom (equivalently, the quantum data associated to a triangle on their boundary).}
\label{fig:gluing}
\end{figure}

The resulting state for the graph $\gamma$ will be a (linear combination of) spin network(s) living in the Hilbert space

\begin{equation}
\mathcal{H}_\gamma = \bigoplus_{\{j\}}\left( \bigotimes_{\{v\}} \mathcal{I^{\vec{j}_{v}}}\bigotimes_{\{e\}\in\partial\gamma}V^{j^e} \right)
\end{equation}
where we considered the general case of a graph including some open (semi-)links $\{e\}$, with all graph links labelled by an irrep of $SU(2)$ and graph vertices labelled by an intertwiner between the associated irreps.

These spin network states constitute the kinematical Hilbert spaces of several quantum gravity formalisms: canonical loop quantum Gravity, spin foam models, simplicial quantum gravity and tensorial group field theories (for quantum geometric models within this broader framework). The difference between these formalisms lies in how the single-graph (or before that, the single-vertex) Hilbert spaces are embedded into the full Hilbert space that includes all possible graphs and all possible numbers of vertices (as necessary to include the infinite number of degrees of freedom one would a priori expect in a full quantum theory of gravity. For example, in group field theory, this can be given by a Fock space of quantum tetrahedra. Generically, in all such formalisms generic states are thus superpositions of (open) spin network states, including a superposition of graph structures.

To appreciate further the role of entanglement in these quantum gravity states, it is interesting to point out a minimal version of entanglement/geometry correspondence in their structure \cite{Colafranceschi:2020ern}.  We have already seen how entanglement is directly encoding graph connectivity (simplicial adjacency relations) and thus the only topological information, in fact, that is encoded in such quantum gravity states, absent any embedding of the graphs inside continuum manifolds. 
Moreover, a local measure of entanglement between simplices glued across a shared face is given by the dimension of the Hilbert space of shared states, i.e. the Hilbert space associated to the  irrep $j$ associated to the dual link: $D = 2 j + 1$; in fact, this is also how the quantum area of the same triangular face scales, upon quantization of the classical area function ("entanglement/area correspondence"). Further, one can ask what is the entanglement between the four triangles/links associated to the same tetrahedron/vertex and the simplest measure is again the dimension of the corresponding Hilbert space of states, which scales like the intertwiner label; in turn, this scales like the quantum volume of the tetrahedron, obtained again by quantizing the corresponding volume function ("entanglement/volume" correspondence).

Before we summarize a few recent results exploiting this entanglement structure, it is also worth emphasizing that such quantum gravity states can be understood in two additional ways that make their quantum information theoretic nature manifest. 

First, they can be understood as generalised tensor networks, of the kind that have been central in much recent literature in quantum many-body physics, entanglement renormalization, numerical simulations of many-body systems, lattice gauge theory, neural networks, quantum computing, AdS/CFT correspondence and more \cite{Orus:2013kga, Verstraete:2008cex, Cirac:2020obd, Singh:2017tet, Tagliacozzo:2014bta, Evenbly:2015ucs}. More precisely, they are generalised Projected Entangled Pair States (PEPS), the generalization corresponding to the fact that the link bond dimension, normally held fixed and equal in all links of the network and her3e corresponding to the dimension of the assigned irreps of $SU(2)$, is dynamical and assigned independently in each link, and to the fact that a generic state is actually a superposition of tensor networks with given combinatorics and bond dimensions, with the superposition affecting also the combinatorial structure (one has a superposition of different graphs).

Second, they can be reformulated as defining quantum circuits \cite{Chen:2021vrc} (see also \cite{Czelusta:2020ryq}). Consider a spin network state associated to an oriented graph with a number of open links, and consider the corresponding spin network wavefunction, depending on group elements (holonomies) associated with the bulk links, in the corresponding irreps of $SU(2)$. This wavefunction can be seen as a boundary-to-boundary map, from the Hilbert space corresponding to the tensor product of representation spaces for the incoming boundary links to the one corresponding to the tensor product of the representation spaces for the outgoing boundary links:

\begin{equation}
\psi(\{ g_e\}_{e\in\gamma}) :  \bigotimes_{e\in\partial\gamma, t(e) \in\gamma} V_{j_e} \rightarrow \bigotimes_{e\in\partial\gamma, s(e) \in\gamma} V_{j_e} \; .
\end{equation}
This map defines a quantum circuit for the degrees of freedom living in the boundary Hilbert spaces, with holonomies playing the role of unitary one-spin gates and intertwiners being instead multi-spin gates. 

\begin{figure}[!ht]
\centering
\includegraphics[width=0.6\linewidth]{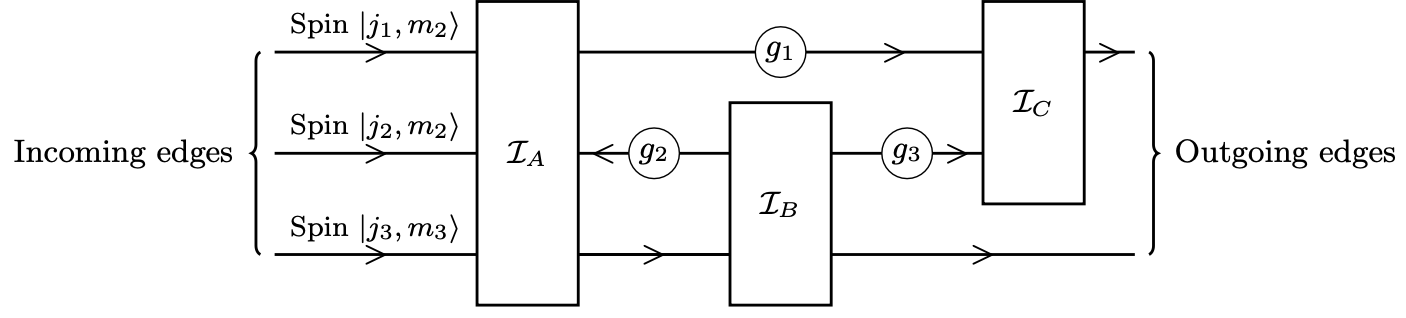}
\caption{\small The quantum circuit corresponding to a spin network wavefunction; holonomies are one-spin unitary gates, while intertwiners are multi-spin gates.}
\end{figure}

This reformulation is intriguing as it is potentially useful for further applications of quantum information ideas in fundamental quantum gravity.

\subsection{Holographic maps and quantum channels from quantum gravity states}

The correspondence between spin network states and generalised tensor networks has been exploited in a number of works, starting from \cite{Singh:2009cd, Singh:2017tet}, and then in \cite{Han:2017uco, Chirco:2017vhs, Chirco:2019dlx} and more recently in \cite{Colafranceschi:2021acz, Chirco:2021chk, Colafranceschi:2022dig}. Here we give a brief summary of the last set of results. For other related results with a similar formal setting and goals, although a slightly different perspective, see also \cite{Akers:2024ixq}.

The starting point is to consider quantum spin network states associated to a generically open graph (which is held fixed in the following), of the general form:

\begin{equation}
| \varphi_\gamma \rangle = \bigoplus _{\{j\}}\sum_{\{n\}}\sum_{\{\iota\}} \varphi^{\{j\}}_{\{n\}, \{\iota\}} \,P_\gamma\, \bigotimes_{v}| \{j^v\},\{n^v\},\iota_v\rangle
\end{equation}
where we have highlighted its construction from a product basis built from the one-body Hilbert spaces, but kept generic the assignment of irreps $j$ for each link and intertwiner labels $\iota$ for each vertex, as well as the vector indices in each representation space. 

\begin{figure}[!ht]
\centering
\includegraphics[width=0.4\linewidth]{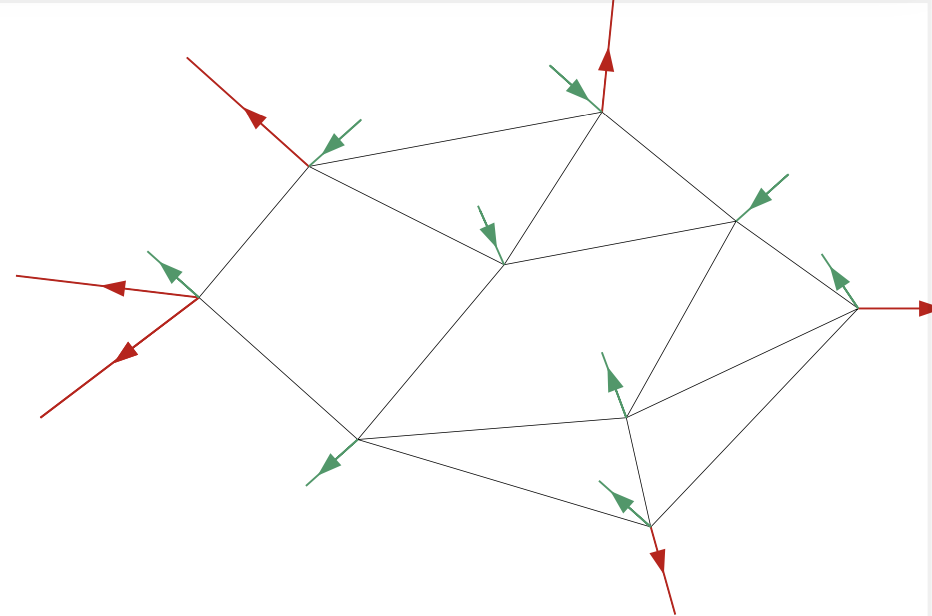}
\caption{\small A schematic representation of a quantum state associated to an open (4-valent) graph.}
\end{figure}

Given such states, one can identify two subsets of data: the spin labels and magnetic numbers associated to the boundary links of the graph, which we may call {\it boundary dofs}, and the spin labels and intertwiner labels associated with the internal links and with the vertices of the graph, respectively, which we call {\it bulk dofs}.

In terms of this partition, we are interested in defining two types of maps for a given quantum state: bulk-boundary maps, and boundary-boundary maps, where the name indicates their domain and target. In the first case, we are interested in particular in the condition that make such bulk-boundary maps {\it holographic}; in the second, we aim also to identify the conditions that would characterize  such boundary-boundary maps as good quantum channels of information.

In order to apply more straightforwardly tensor network techniques, we restrict to a subclass of quantum states, whose modes factorise per vertex, i.e. $\varphi^{\{j\}}_{\{n\} \{\iota\}} = \prod_{v}(f_v)^{j^v}_{n^v, \iota_v}$.

Consider the first issue.
To simplify the analysis, we fix the bulk spins to specific values collectively labelled $J$. This leads to a decoupling of boundary and remaining bulk dofs. Thus the relevant Hilbert space factorizes as:  $\mathcal{H}^{J} = \bigotimes_v \mathcal{I}^{\vec{j}_v} \bigotimes_{e \in \partial\gamma}V^{j_e}$. Now we can define a map $M_{\varphi_\gamma}$, depending on the chosen quantum state, between the two bulk and boundary sub-spaces, mapping a generic bulk state $| \zeta \rangle = \sum{\{ \iota\}}\zeta_{\{\iota\}}| \{\iota\}$ to the boundary state $| \varphi_{\partial\gamma}(\zeta)\rangle = \langle \zeta | \varphi_\gamma\rangle$, which is clearly fully specified by the chosen quantum state for bulk$+$ boundary.

As a proxy condition for holographic behaviour, we take the isometry of this map, i.e. the condition: $M_{\varphi_\gamma}^\dagger M_{\varphi_\gamma} = I$, where $I$ is the identity in the bulk subspace.

One can then show that the isometry condition is satisfied if and only if the reduced bulk density matrix $\rho_{bulk} = Tr_{\partial\gamma}\left[\frac{|\varphi_\gamma\rangle\langle \varphi_\gamma|}{\prod_v D_{j^v}}\right]$ is maximally mixed, i.e. it has maximal entropy. In turn its entropy can be estimated, in terms of Renyi entropies, via standard randomization techniques applied to tensor networks \cite{Hayden:2016cfa, Qi:2017ohu}, which allow to translate the problem of maximizing the entropy of the reduced density matrix of the quantum system into that of minimizing the free energy of a dual Ising model. In our case, the same randomization method shows that, in the regime in which spins are large (naively, a semiclassical regime),  the boundary-bulk map defined by our quantum gravity state is, roughly speaking, the more isometric (holographic) the more inhomogeneous is the assigment of spin labels. The precise mathematical conditions can be found in \cite{Colafranceschi:2021acz}. For related work, although relying on different methods, see \cite{Chen:2021vrc}. This is interesting, because it may indicate an avenue for a microscopic, quantum gravity realization (and explanation?) of holography in a non-spatiotemporal and information-theoretic context.

Consider now the second issue, i.e. transmisison of information and the entanglement between two portions of the boundary, for the same quantum state and within the same restrictions (fixed bulk spins, fixed graph, factorized state).  
We partition the boundary dofs into two complementary sets $A$ and $\bar{A}$, and look at the reduced density $\rho_A = Tr_{\bar{A}}[\rho]$ for the region $A$, where $\rho$ is the density matrix for the full quantum state $\varphi_\gamma$. 

\begin{figure}[!ht]
\centering
\includegraphics[width=0.4\linewidth]{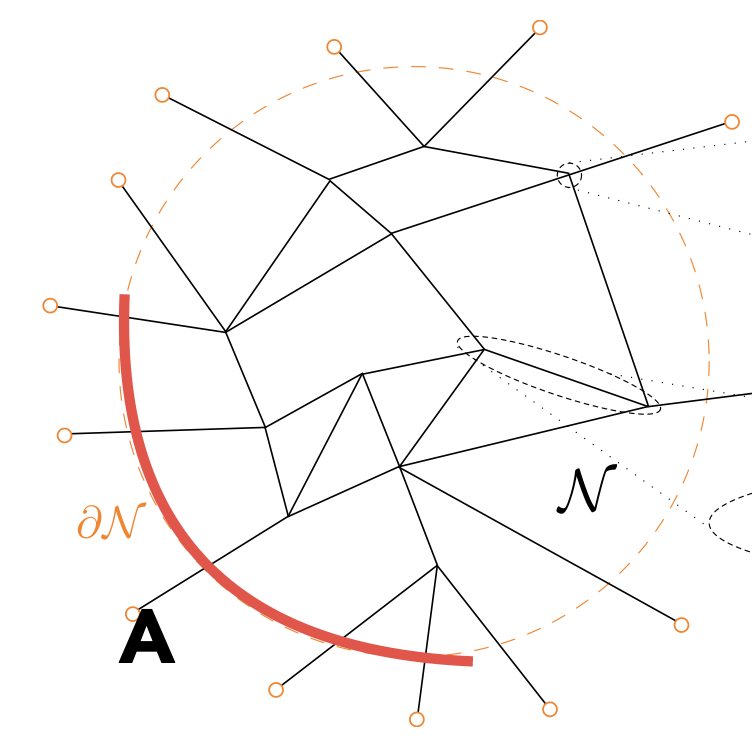}
\caption{\small Partitioning the boundary links into two subsets}
\end{figure}

We are interested in the entanglement entropy between the two subregions of the boundary. As a proxy for it, we can compute the 2nd Renyi entropy of the reduced density matrix, using again the same random tensor network techniques, and thus the same dual ising model. The calculation can be performed for both the homogeneous, same-spin case (all bulk spins assumed equal), and the inhomogeneous one. 

From the calculation, in the case of vanishing bulk (intertwiner) entropy, one obtains an exact Ryu-Takayanagi-like formula

\begin{equation}
S(\rho_A)_2 \simeq K(j,\gamma) \text{min}_{\Sigma_A}| \Sigma_A| 
\end{equation}

where $K$ is a factor depending on the details of the bulk spin assignment and $| \Sigma_A|$ is the size (i.e. the number of crossing links) of the minimal surface in the bulk separating the two boundary regions.

\begin{figure}[!ht]
\centering
\includegraphics[width=0.4\linewidth]{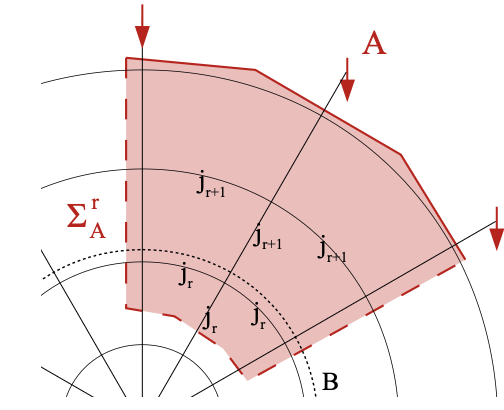}
\caption{\small RT-like behaviour for the entanglement entropy between boundary subregions}
\end{figure}

When the bulk (intertwiner) entropy is not negligible, one finds that the RT formula acquires a correction term measuring the bulk contribution 
\begin{equation}
S(\rho_A)_2 \simeq K(j,\gamma) \text{min}_{\Sigma_A}| \Sigma_A| \, +\, S_{bulk}
\end{equation}

which can also be computed. Interestingly, when the bulk entropy increases, a smaller and smaller portion of the RT surface enters the bulk regions, and, under the same increase, when the boundary region $A$ tends to occupy the whole boundary, the RT surface tends to coincide with the boundary of this high-entropy bulk region. This is closely reminiscent of a black hole horizon, whose surface coincides with the RT surface, in the continuum geometry picture, and which encloses a maximal entropy bulk region. This is intriguing because it suggest a possible realization of holographic behaviour and of effective black hole geometries in a non-spatiotemporal and information-theoretic context.

\begin{figure}[!ht]
\centering
\includegraphics[width=0.4\linewidth]{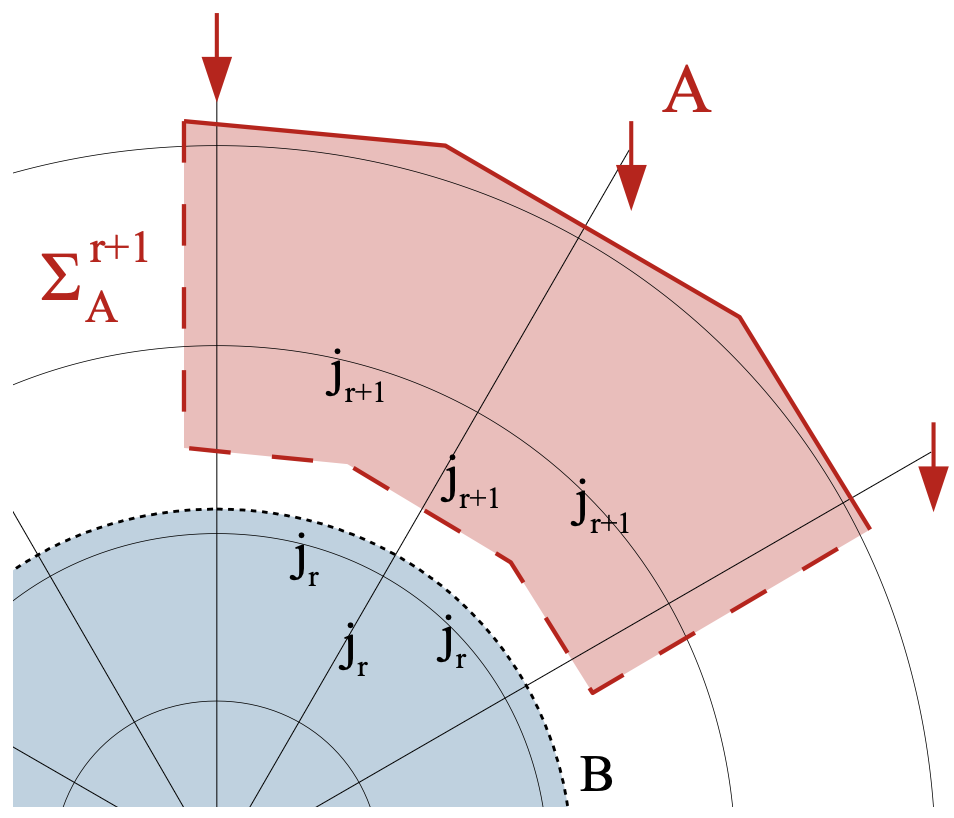}
\caption{\small Black hole-like behaviour for the entanglement entropy between boundary subregions, when bulk entropy grows and the boundary region is extended}
\end{figure}

In trying to perform the same type of analysis for quantum gravity states involving a sum over spins, thus a sum over (discrete) quantum geometries, one has to face the issue that the corresponding Hilbert space does not present any obvious factorization between bulk and boundary dofs or between subsets of boundary ones (but it possesses a direct sum structure with respect to the possible spin assignments to the whole graph). The very notion of entanglement between any subset of dofs becomes ambiguous. If one tries to bypass this ambiguity by embedding the Hilbert space into a larger, factorized one and work with a more clearly defined notion of subspaces and entanglement there, the ambiguity presents itself again in the choice of embedding. One way to proceed is to work at the level of algebras of observables (acting on the given Hilbert space of quantum gravity states) and to identify a notion of subsystem in terms of subalgebras, rather than subspaces. Moreover, holographic behaviour is then best characterized in terms of \lq transmisison of information\rq, rather than entanglement scaling. This strategy has been applied for quantum gravity states in \cite{Colafranceschi:2022dig}, and outlined in its general formal aspects in \cite{Langenscheidt:2024bce}.

Here we just outline the main steps in the analysis. 

Given the full algebra, one can identify algebraic subsystems, i.e. two subsets of observables separately \lq testing\rq the two subsets of quantum dofs one is interested in mapping, to be considered as \lq input\rq and \lq output\rq (it could be bulk and boundary or two portions of the boundary). These are not subalgebras, in general. One can then define a map between input-output operator spaces via the Choi-Jamiolkowski isomorphism. 

The isometry condition on such map, our proxy for holography or complete transmission of information, can be shown to follow from a certain set of necessary condition, corresponding to the requirement that the operator map defines a quantum channel. 

These conditions can be identified using the same random tensor network techniques, in the regime of large spins (which minimizes key quantum fluctuations), and translated again also in conditions on the entropy of the underlying quantum gravity state. 

One obtains again a Ruy-Takanayagi-like formula for the Renyi entropy of the quantum state depending on a set of minimal surfaces, one for each spin sector, thus one for each superposed quantum geometry, and with a definition of \lq area spectrum\rq for the minimal surfaces that can be related to but differs from the one used in canonical loop quantum gravity or simplicial quantum geometry. 

In general, the necessary conditions for isometry amount to have negligible correlations between boundary data and intertwiner bulk data, and on specific peaking properties of the quantum state around  a subset of spin sectors. These same conditions may actually become also sufficient ones, upon additional constraints on the quantum states. 

To summarize, for both boundary-bulk and boundary-boundary maps, holographic behaviour appears to require the bulk Hilbert space to be comparatively small with respect to the boundary one (as measured by the dimensionality of their respective subspaces), and the total boundary area (scaling with the size of spin spaces assigned to its links) to be approximately constant across different subsectors of spin assignments.

Again, the importance of these results is not so much in the details of their conclusions, but in the very fact that (quantum) geometric properties and quantum information theoretic properties are deeply intertwined and can be studied also in a non-spatiotemporal, purely combinatorial and algebraic context, hopefully shedding light on the emergence of holographic (and gravitational) behaviour at macroscopic scales as well as on its fundamental origin.

\section{Quantum gravity processes as quantum causal histories (or not)}

We now turn to the quantum dynamics of the quantum gravity structures we considered in the previous section, and that we characterized in quantum information-theoretic terms. Our main point is that a similar quantum information-theoretic characterization can be provided also for the quantum gravity processes they are subject to, and that information theoretic tools can be applied to the analysis of their (quantum) causal properties. As in the previous section, we consider a subset of quantum gravity formalisms, sharing many of their constitutive structures, and focus on their shared elements, rather than their differences.

\subsection{Quantum causal processes of atoms of space - quantum spacetime as a quantum circuit}
A general scheme for quantum processes respecting minimal causality conditions, and to which we can try to fit or adapt fundamental dynamical processes for our \lq atoms pf quantum space\rq, is represented by the formalism of quantum causal histories. The version we refer to here is the one in \cite{Markopoulou:1999cz, Hawkins:2003vc}, with its initial development in a quantum gravity context to be found in \cite{Markopoulou:1997wi, Markopoulou:1997hu}.\footnote{Of course, the abstract characterization of quantum dynamics and quantum causality, as well as its generalization to a context in which geometry and thus causal relations are themselves dynamical and subject to superposition, is a hot topic in quantum foundations, with many recent developments. See \cite{Oreshkov:2011er, Goswami:2018rda, Baumann:2021urf, mrini2024indefinitecausalstructurecausal, hardy2005probabilitytheoriesdynamiccausal} for a small sample. Obviously, we are not going to review any of that.}  

Possible dynamical processes are given by a set of \lq events\rq together with an order relation between pairs of them; these are also the constitutive elements of a directed graph. In 4d quantum gravity models based on (quantum) simplicial geometry, fundamental events may be taken to correspond to 4-simplices, while order relations between pairs of them correspond to their shared 3-simplices. The directed graph would then correspond to the dual 1-skeleton of the oriented simplicial 4-complex. Note that this realization implies a restriction to 5-valent directed graphs. For Lorentzian models, the order relations can be given a causal interpretation. An important special case is represented by partially ordered sets (posets), which are directed graphs which are also irreflexive, i.e. do not contain closed causal loops. Posets are also called, in the quantum gravity literature, {\it causal sets} and are the basic entities in the causal set approach to qauntum gravity \cite{Dowker:2024fwa}.

\begin{figure}[!ht]
\centering
\includegraphics[width=0.5\linewidth]{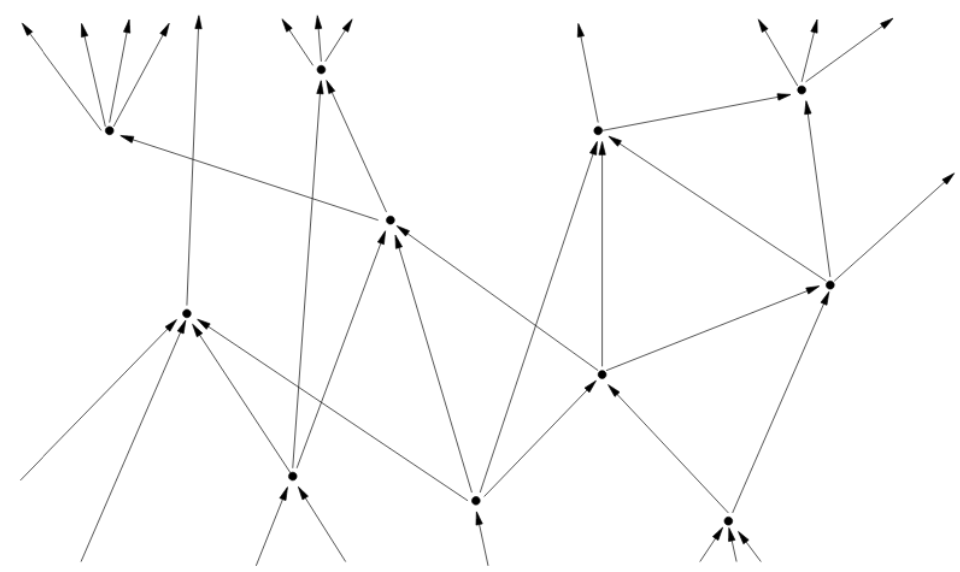}
\caption{\small An example of a 5-valent directed graph}
\end{figure}

Clearly, this structure can be decomposed into elementary \lq evolution steps\rq, corresponding to the possible orientation assignments of the 5-valent nodes: 5 links outgoing, 1 link incoming/4 links outgoing, 2 links incoming/3 links outgoing, and their inverses.

The quantum process corresponding to each directed graph is obtained by an assignment of Hilbert spaces to the links (and tensor products of Hilbert spaces for unordered sets of links) and elementary \lq evolution\rq operators to the nodes; in addition one can include also \lq gluing operators\rq to the links, enforcing a prescription for the \lq transmission of information\rq ~from one node/event to another \footnote{In fact, a more complete and consistent definition of a quantum process in this language, of of a quantum causal history in particular, is given in terms of an assignment of algebras of operators and completely positive maps \cite{Hawkins:2003vc}. We give here a simplified earlier construction, which is sufficiently indicative of the general points we want to make.}.

\begin{figure}[!ht]
\centering
\includegraphics[width=0.5\linewidth]{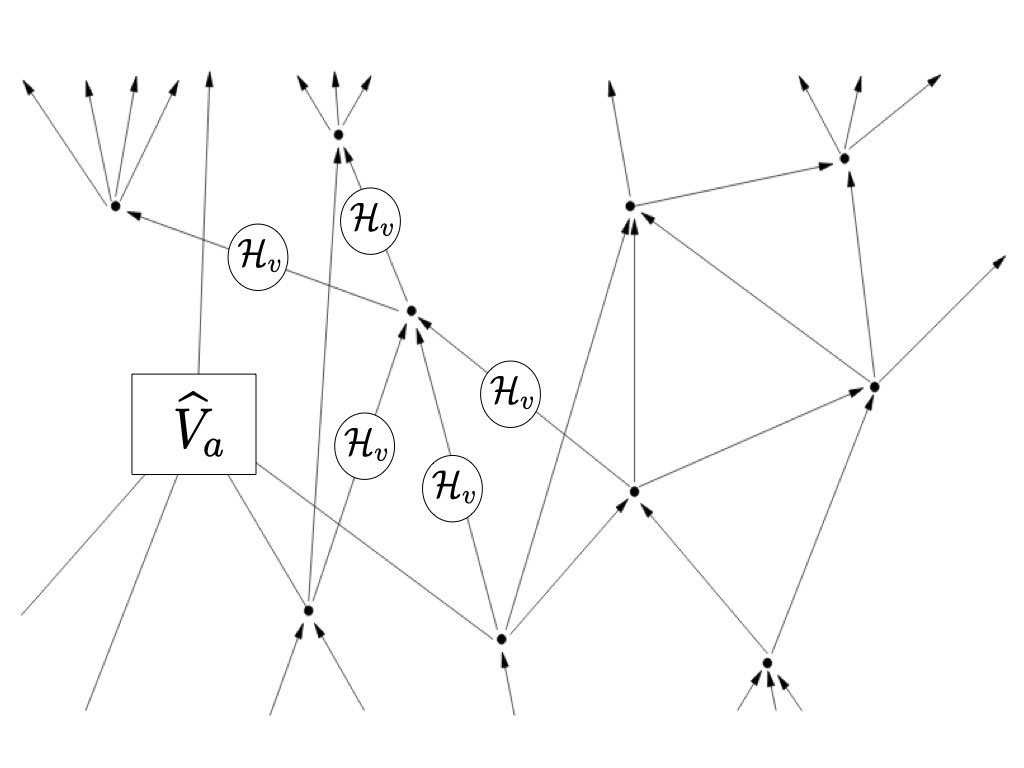}
\caption{\small An example of an elementary quantum process}
\end{figure}

A specification of a quantum dynamics and thus a specific quantum gravity model corresponds to the assignment of a quantum (probability) amplitude to each quantum process (together with any restriction on the class of allowed processes, be it on the combinatorial structure or the associated operators and Hilbert spaces). Such amplitude is defined by the chosen node operator and gluing operator, in turn characterized by the corresponding kernels

\begin{eqnarray}
\mathcal{V}_n : \underset{e \in \partial n}{\otimes} \mathcal{H}_v^e \longrightarrow \mathbb{C} \nonumber \\  \mathcal{P}_e : \mathcal{H}_v^e \otimes \mathcal{H}_v^{e *} \longrightarrow \mathbb{C} \nonumber
\end{eqnarray}
where we have left implicit the dualization of the Hilbert spaces required to reflect the different orientations of the links associated to the nodes. Notice that the specification of a gluing operator requires a \lq doubling\rq of the Hilbert space associated to the links of the directed graph, distinguishing one copy of it associated to one of the connected nodes from the (dualized) one associated to the other connected node\footnote{Notice that this doubling is necessary, if one wants to allow for a composition of processes or, from the point of view of their dual cellular complexes, the composition of different cellular complexes along shared boundaries.}. Given these building blocks, the quantum amplitude associated to the given process $\Gamma$ can be defined as:

\begin{equation}
\mathcal{A}(\Gamma) = \underset{e\in\Gamma}{Tr}\left( \prod_{e\in\Gamma} \mathcal{P}_e \prod_{n\in\Gamma}\mathcal{V}_n\right)
\end{equation}

where with $Tr$ we have indicated the trace operation over the (doubled) Hilbert spaces associated to the links of the process. Different quantum gravity formalisms (notably canonical loop quantum gravity, spin foam models, lattice gravity path integrals and group field theories) share this general structure of their dynamical quantum amplitudes, in a covariant language. For example, for a more detailed presentation in the case of spin foam models (adapted also to the group field theory language) see \cite{Finocchiaro:2018hks}, in addition to the existing introductions to all these quantum gravity formalisms.

In fact, in quantum gravity formalisms that use this kind of discrete structures, the complete quantum dynamics should be completed with the definition of a continuum limit/approximation, since reliance on any given cellular complex represents at best a truncation of the full set of quantum degrees of freedom of the fundamental theory. Also in accordance with the superposition principle and the interpretation of the above discrete structures as possible quantum processes, this limit is encoded within a sum over all complexes/processes within the allowed class. The full quantum dynamics is therefore encoded in a partition function (or transition amplitude, if the complexes have boundaries) of the general form:

\begin{equation}
Z = \sum_{\Gamma} w(\Gamma) \mathcal{A}(\Gamma)
\end{equation}
where again the specification of the additional weights $w(\Gamma)$ concurs to the definition of the specific model being considered.

Two points are worth emphasizing about this construction. First, also these quantum processes, like the quantum states discussed in the previous section, can be reformulated in the language of tensor networks, and this has been applied to the study of renormalization of quantum gravity dynamics from a lattice gauge theory perspective \cite{Dittrich:2014mxa}. Second, at this level, thus before any extraction of coarse-grained or otherwise effective continuum dynamics, the connection of these dynamical quantum amplitudes to that of gravitational theories can only be with lattice gravity, i.e. some discretized version of gravitational physics. This connection depends strongly on the specific quantum gravity formalism being considered. In spin foam models, thus also in group field theories, the connection can be shown (modulo remaining technical open issues) in the form of an exact equivalence with lattice gravity path integrals for 1st order gravity theories\footnote{Usually, with gravity formulated as a constrained topological theory \cite{Bianchi:2012nk}. But see \cite{Baratin:2008du, Dekhil:2025dma} for models constructed following alternative strategies.} \cite{Finocchiaro:2018hks}, or in a semi-classical, asymptotic regime with 2nd order gravity \cite{Asante:2020qpa} (in terms of area variables \cite{Dittrich:2023ava}).

\

We are interested in identifying the requirements on such quantum processes that make them {\it bona fide} quantum causal histories, that is, that make each quantum process encore a properly causal unitary evolution. We do not discuss here to what extent this is something we should expect or require, from fundamental quantum gravity processes, besides a few short remarks.

The required properties have been identified in \cite{Markopoulou:1999cz, Hawkins:2003vc}. Consider the evolution operator $E_{\alpha\beta} : \mathcal{H}_\alpha \rightarrow \mathcal{H}_\beta$, obtained by composing all the elementary operators $V$ connecting two complete a-causal subsets $\alpha$ and $\beta$ of links, that is subsets of links that are all causally unrelated within each subset and one the complete causal future (past) of the other, where the Hilbert spaces $\mathcal{H}_{\alpha,\beta}$ are the tensor product of the Hilbert spaces associated to the constituting links.
For the quantum process to define a quantum causal history, the operators $E_{\alpha\beta}$  should be: a) reflexive: $E_{\alpha\alpha} = I_\alpha$, b) antisymmetric: $E_{\alpha\beta}E_{\beta\alpha} = I_\alpha \Leftrightarrow E_{\alpha\beta} = E_{\beta\alpha} = I_\alpha$; c) transitive: $E_{\alpha\beta}E_{\beta\gamma} = E_{\alpha\gamma}$; d) unitary: $\sum_{\beta} E_{\alpha\beta}E_{\alpha\beta}^{\dagger} = \sum_\beta E_{\alpha\beta}\bar{E}_{\beta\alpha} = I_\alpha$.

One can then check or impose that the relevant evolution operators of quantum gravity models of interest satisfy these requirements. However, one could also question their necessity. Indeed, we have stated that the full quantum gravity dynamics should rather be given by a sum over all such quantum processes. Therefore, one could argue that the operators satisfying proper causality conditions are those obtained by performing this sum, that is:

\begin{equation}
\mathcal{E}_{\alpha\beta} = \sum_c \lambda_c E^c_{\alpha\beta} \, :\, \mathcal{H}_\alpha \longrightarrow \mathcal{H}_\beta 
\end{equation}
between the same complete a-causal subsets.
In other words, even accepting that the fundamental quantum gravity dynamics should be expressed in the language of quantum causal histories, one could argue that it is the full transition amplitudes that should satisfy causality constraints, and not the possible individual quantum processes.

It is easy to see \cite{Livine:2002rh} that these two perspectives are not equivalent, and in fact, they are not even consistent with one another. More precisely, while imposing \lq micro-reflexivity\rq (i.e. reflexivity of individual processes) implies full reflexivity (i.e. reflexivity of the evolution obtained upon summing over micro-processes) and the same is true for antisymmetry, the other two requirements are much more problematic. First, one could make micro-transitivity compatible with transitivity of the full evolution, by defining the latter more generally as $\sum_{\beta}\mathcal{E}_{\alpha\beta}\mathcal{E}_{\beta\gamma} = \mathcal{E}_{\alpha\gamma}$, which is just the standard composition of quantum (transition) probability amplitudes; but micro-transitivity itself appears to be too strong a requirement from the discrete quantum gravity perspective; indeed, taking into account also the dual lattice formulation of quantum processes, micro-transitivity is equivalent to a condition of partial triangulation invariance of the quantum dynamics, i.e. a requirement that the quantum amplitudes associated to individual lattices are partially independent of the chosen lattice. To the extent in which gravity is not a topological field theory with no local propagating degrees of freedom, this is a suspicious condition since it may imply a partial triviality of the quantum dynamics. Second, one can actually verify that unitarity of the full evolution implies that the micro-evolution {\it must not be unitary}. So one has to make a choice.

It remains a valid goal to have the full quantum gravity dynamics, obtained via a sum over elementary quantum processes, defining a (\lq coarse-grained\rq) quantum causal history. 
This would be interesting, from the perspective we are exploring in this contribution, because it would mean that quantum information and quantum computation can be found at the very heart of quantum gravity also at the dynamical level. The same interest would remain if instead one insists on the elementary processes be themselves quantum causal histories.

Indeed, it is a general result \cite{Livine:2006xc} that a quantum causal history admits a unitary evolution between its acausal surfaces if and only if it can be represented as a quantum computational network, i.e. a quantum circuit.

The idea of spacetime as a quantum circuit would then find a concrete realization, if quantum gravity evolution can be formulated as in terms of quantum causal histories, at the elementary or coarse-grained level, in the above language or in the more refined one of observable algebras and CP maps \cite{Hawkins:2003vc}. See also \cite{Arrighi:2016zod, Arrighi:2021xsn} and the cited literature on process matrix formalism and indefinite causal structures for related directions.

Let us now look in more detail at a couple of features of quantum gravity processes that have to be realized in order for them to define quantum causal histories, thus quantum circuits, and see which obstacles one has to face to do so.

\subsection{Causal hiccups and causal indifference in QG processes}
As discussed, a proper representation in terms of quantum circuits requires: a) the absence or \lq\lq irrelevance\rq\rq of  causal loops; b) suitable conditions ensuring unitarity of the evolution operators. 

\

Concerning closed causal loops, there are three possible strategies that can be followed in constructing quantum gravity models: a) define a quantum dynamics (amplitudes) that eliminates causal loops altogether; b) define a quantum dynamics (amplitudes) that suppresses causal loops, by assigning them subdominant contributions, in the relevant regimes; for example, one could consider admitting causal loops in the theory, provided they do not spoil expected semi-classical or continuum physics; c) define a quantum dynamics (amplitudes) that only allows \lq\lq harmless\rq\rq. Let us stress that the directed graphs underlying all current spin foam models and lattice gravity path integrals (or group field theory perturbative amplitudes) contain closed loops of order relations, i.e. causal loops. Thus the issue is of concrete relevance for such quantum gravity formalisms.
While the first two options may be technically very challenging but are conceptually straightforward, when exactly a causal loop is physically harmless requires a more careful analysis. The issue has been studied for example in \cite{Livine:2006xc}, in the quantum gravity context, following earlier work by Deutsch \cite{Deutsch:1991nm} in a general quantum mechanical context. The upshot is that causal loops are either entirely disruptive or entirely harmless, to paraphrase. More precisely, they either prevent the standard formulation of quantum mechanics to be applicable or, when they do not, they lead to no observable changes, since they simply end up contributing an extra subspace to the ordinary causality-respecting system. Moreover, by applying suitable (Deutsch) criteria, the causally well-behaved region of the process decouples entirely from the causal loop, if the quantum dynamics remains linear, and if it does not stay linear, then the causal loop does not carry independent degrees of freedom. For more details we refer again to \cite{Livine:2006xc}.

\

Let us now discuss unitarity of quantum evolution and, before that, the very dependence of the quantum gravity transition amplitudes from the order of their arguments, thus a notion of past/future relating the quantum states it depends on, which is in many ways a prerequisite for it. The issue is: given two (\lq initial\rq and \lq final\rq) quantum states, which kind of quantum amplitude do we define, via the gravitational path integral? Let us first recall a few facts about quantum gravity path integrals, which can be verified at the formal level in great generality \cite{Teitelboim:1981ua, Halliwell:1990qr}, and have to be then realized concretely in more rigorous manner by quantum gravity approaches, including the ones based on discrete structures that we have focused on here. Different quantum gravity amplitudes can be defined starting from the same proper-time truncation of the full path integral (in canonical form), obtained via appropriate gauge-fixing of the general expression:
\begin{equation}
K[h_{ij}^2, h_{ij}^1; N(\tau_2 - \tau_1)] = \int \mathcal{D}h_{ij}(x,\tau)\mathcal{D}\pi^{ij}(x,\tau) e^{i S(h_{ij},\pi^{ij}); N}
\end{equation}
where the amplitude depends on the fixed metric data on the two (past/future) boundaries. This expression could be required to corresponds to (the matrix elements of) a unitary evolution operator in proper time. 
However, this is not the physical (transition) amplitude for quantum gravity, since the lack of integration over proper time (lapse) means that we have not yet imposed any conditions encoded in the Hamiltonian constraint of the (canonical) theory, thus no full quantum gravitational dynamics (the Einstein's equations are indeed encoded in the Hamiltonian constraint), beside enforcing at best some semiclassical restriction (due to the appearance of the gravitational action in the expression). 
One can then define a \lq causal\rq transition amplitude (the analogue for quantum gravity of what would be, for a relativistic particle, the Feynman propagator, by integrating the lapse (proper time) over the full positive range:

\begin{equation}
\mathcal{K}[h_{ij}^2, h_{ij}^1]= \, \int^{N(x)= +\infty}_{N(x)= 0} \mathcal{D}[N(x)(\tau_2 - \tau_1)]\, K[h_{ij}^2, h_{ij}^1; N(\tau_2 - \tau_1)] \, . 
\end{equation}

This is the canonical counterpart of the straightforward Lagrangian gravitational path integral $\mathcal{K}[h_{ij}^2, h_{ij}^1]= \, \int\mathcal{D}g_{\mu\nu} e^{i S(g_{\mu\nu})}$. it is invariant under Lagrangian (covariant) diffeomorphisms and indeed switches to its own complex conjugate under switch of spacetime orientation; in other words, it does register an ordering between its two arguments, the boundary quantum states. It does not, however, give a solution to the canonical Hamiltonian constraint, i.e. it is not invariant under canonical symmetries (the canonical Dirac algebra, counterpart of covariant diffeomorphisms, which are a subset of the canonical ones). A solution of the canonical Hamiltonian constraint is instead obtained by integrating the lapse (proper time) over the full (positive and negative) real values:

\begin{equation}
\mathcal{C}[h_{ij}^2, h_{ij}^1]= \, \int^{N(x)= +\infty}_{N(x)= -\infty} \mathcal{D}[N(x)(\tau_2 - \tau_1)]\, K[h_{ij}^2, h_{ij}^1; N(\tau_2 - \tau_1)] \, . 
\end{equation}
This indeed defines (formally)  a physical scalar product between canonical quantum gravity states, solving all the constraints of the theory (or, equivalently, the matrix elements of the projector operator onto such solutions). Its Lagrangian counterpart would look like $\mathcal{C}[h_{ij}^2, h_{ij}^1]= \, \int\mathcal{D}g_{\mu\nu} \left[ e^{i S(g_{\mu\nu})} \, +\, e^{-iS(g_{\mu\nu})}\right]= \int\mathcal{D}g_{\mu\nu} \; \cos(S(g_{\mu\nu}))$ . This quantity does {\it not} register the spacetime orientation and it is symmetric under its switch, not encoding any ordering among it arguments, the \lq initial\rq and \lq final\rq quantum states.

Spin foam models (equivalently, the perturbative transition amplitudes of group field theories) aim to be discretized and thus mathematically better defined realization of the gravitational path integral. Which of the above quantities do they actually realize?

All the most studied spin foam models are discrete counterparts of the path integral for gravity formulated as a constrained topological BF theory. It turns out that, like their continuum counterpart (and the path integral for topological BF theory itself) they are invariant under switch of spacetime (lattice) orientation, more precisely the inversion of the orientation of their constitutive simplicial structures; this invariance is in fact realized locally at the level of each node or 4-simplex contribution $\mathcal{V}$ to the total amplitude $\mathcal{A}$ as well as at the level of lower-dimensional simplices (eg triangles) in the simplicial complex. Recall that the 1-skeleton of this simplicial complex corresponds to the directed graph whose order relations have a tentative causal interpretation (in Lorentzian models). The orientation independence of the spin foam amplitudes thus implies than none of the most studied spin foam models defines a proper quantum causal history (a quantum circuit) and a unitary quantum gravity dynamics. This would remain true even if one was able to remove causal loops from the underlying directed graph.

It is possible to construct \lq properly causal\rq spin foam models, but a suitable restriction of their amplitudes so that they register faithfully the orientation of the underlying complex. This has been done first, for the Barrett-Crane model, in \cite{Livine:2002rh}, and more recently by similar procedures for the EPRL model in \cite{Engle:2015mra, Bianchi:2021ric}. These restricted models are therefore candidates for the realization of the \lq causal propagator\rq for quantum gravity, and for a formulation in terms of quantum causal histories and quantum circuits. However, these causality-inspired constructions are all rather {\it ad hoc} and we still lack a systematic construction procedure of spin foam models from first principle (rather than by restricting by hand a-causal models) taking into account causality restrictions, as well as a more complete analysis of the properties of the present \lq ad-hoc\rq ones.

\section*{Conclusions}
We have argued that both semi-classical considerations and quantum gravity formalisms suggest, in different ways, that spacetime and gravity may be emergent, collective, not fundamental notions, and that the universe may be a (peculiar, background independent) quantum many-body system of pre-geometric quantum entities, some yet to be unraveled \lq atoms of space\rq . In particular, there are intriguing indications that topology and geometry may emerge from the entanglement among such fundamental quantum entities. More generally, an intriguing possibility is that quantum spacetime physics may be formulated, in its most fundamental level, entirely in the language of quantum information.

A variety of quantum gravity formalisms share the same combinatorial and algebraic quantum structures as quantum states:
quantized simplicial structures and spin networks. We have outlined the ways in which such quantum states can be described in quantum information-theoretic terms. More precisely, we have summarized how these quantum states: a) can be seen as generalised tensor networks and realize a precise discrete entanglement/geometry (and topology) correspondence; b) can be framed as information channels (or quantum circuits); c) can be used to define bulk/boundary and boundary-to-boundary maps, for which one can then identify conditions for holographic behaviour. This could indicate an avenue toward understanding the microscopic origin of holographic behaviour in quantum gravity.

We have then discussed how, in the same quantum gravity formalisms, dynamical quantum processes can be recast as quantum causal histories, provided some key properties are implemented in their amplitudes, and then again as quantum circuits; we have also pointed out some of the challenges faced to implement the required properties Again, the main point is that quantum  information tools and language may be the appropriate ones to formulate also the quantum dynamics of the microscopic constituents of the universe, when geometry and fields fail.

Before we go on to comment on some more conceptual aspects of the these conclusions, we point out that the (tensorial) group field theory framework, on top of providing a completion of lattice gravity path integrals and spin foam models (thus sharing the same quantum amplitudes) and a 2nd quantized framework for spin network states (thus a convenient Fock space structure for their Hilbert space) \cite{Oriti:2011jm, Oriti:2014uga}, provides also a number of almost standard field theoretic tools to study them and in particular to extract effective continuum gravitational physics from them \cite{Oriti:2021oux}. This means that, even in this more abstract, non-spatiotemporal, pre-geometric context, one can apply quantum field theoretic techniques to the quantum information structures we presented above, to analyse their formal properties and to unravel their physical meaning.  

\

\noindent {\bf Some philosophical considerations: is the universe a quantum computer?} -  To conclude let us offer some thoughts on different ways in which we could interpret the relevance of quantum information language and tools for encoding the fundamental microstructure of spacetime and the universe. 

A straightforward attitude is to give to this fact an ontological basis: the universe is a quantum computer. In this case, the fact that quantum information is the appropriate language to understanding is no surprise: it is just the language representing how it fundamentally operates, it is the language that constitutes its basic laws. From the epistemological point of view, this attitude follows from and it is grounded on a straightforward scientific realism: the world is out there and entirely independent, in its properties, of our epistemic activities, which achieve (at best) a faithful (albeit partial) representation of the way the world is. It is also tied to a realist and ontologically committed view on laws of nature: they are what governs the physical world, i.e. the rules by which it functions and evolves. 

Challenges against all the above are numerous and the philosophical debate about each of the above points is old and intricate and interesting. Here we want to make two brief comments about this \lq the universe is a quantum computer\rq view, implicitly based on the attitude we just summarized.

A first immediate one is that all the realist views on laws face challenges at different levels from quantum gravity, and in particular from the very possibility of space, time and geometry being emergent notions. The quantum information structures we discussed in this contributions can be seen as encoding these challenges, but at the same time offering tentative ways to meet (if not solve) them. This is discussed in \cite{Lam:2024ibm}.

A second one is that one can adopt (and try to develop further) a less ontologically committed view on the object of our scientific theories, i.e. the physical world, including laws of nature, and thus a weaker version of (scientific) realism. One can adopt a more epistemic view on physical laws based on a more substantial role given to epistemic agents, taken to be irreducible and not negligible (outside convenient idealizations). This epistemic, agent-based view would tie well with a more participatory form of realism, in which what is real is, roughly speaking, only the result and content of the interaction between the world and the epistemic agents, none of which is independently real outside of such interaction. This weaker form of realism would be the only kind of ontological commitment allowed by the epistemic premises.  This view may have implications for (and be tested with) the interpretation of quantum mechanics, as well as the construction and interpretation of theories of quantum spacetime and geometry. 

More generally, it would lead to a view in which the universe is (to a large extent) what we think it is (or what we model it as), in the sense that it is our epistemic constructions that {\it make reality}, rather than simply represent it. Obviously this is just a vague statement, to be made more precise and articulated, but it is maybe interesting to see how a similar view changes how we may interpret the role of (quantum) information theoretic structures in fundamental quantum gravity. In a fundamental quantum gravity context, we argued, we have no spacetime notions to rely on; we have to think the world (and model it) without spacetime. We are then left with combinatorics and algebra as mathematical language, and with information processing, rather than definite, ontologically grounded objects (whose ontological characterization would normally {\it assume} space and time), as the only \lq dynamical\rq content. This reflects the more basic, more irreducible structures in our thinking, which at the same time correspond (given the above view of what it is to be \lq real\rq) to the more basic structures \lq in the world\rq. (Quantum) Computers are abstract models of (quantum) information processing, and of our own thinking. To the extent in which the universe is (largely) what we think it is (in the sense outlined above), and we think like (quantum) computers, it is not so surprising, perhaps, that the quantum (non-spatiotemporal) universe is naturally modeled as a quantum computer.

These rather vague considerations are offered, here, only as potentially useful indications of philosophical avenues to explore and develop further, on the basis of the scientific developments in quantum gravity, that we have summarized in this contribution.

\section*{Ackowledgements}
We acknowledge  support through the Grant PR28/23 ATR2023-145735 (funded by MCIN /AEI /10.13039/501100011033). We also thank the organizers and the participants of the \lq\lq QG and computation\rq\rq workshop in Sydney for a very interesting event and for many stimulating discussions, as well as the editors of this volume for their patience.

\newpage 

\bibliography{references}
\bibliographystyle{jhep} 
\end{document}